\documentclass[twocolumn,showpacs,aps,prl,amsmath,amssymb,floatfix,%
superscriptaddress]{revtex4}
\usepackage[dvips]{color}
\usepackage{graphicx}
\usepackage{epsfig}
\usepackage{dcolumn}
\usepackage{bm}
\begin{document}
\title{Kondo Shuttling in Nanoelectromechanical Single-Electron Transistor}
\author{M.N. Kiselev}
\affiliation{Institute f\"ur Theoretische Physik, Universit\"at
W\"urzburg, 97074 W\"urzburg, Germany} \affiliation{Material
Science Division, Argonne National Laboratory, Argonne, Illinois
60439}
\author{K. Kikoin}
\affiliation{Physics Department, Ben-Gurion University of the
Negev, Beer-Sheva 84105, Israel}
\author{R.I. Shekhter}
\affiliation{Department of Physics, G\"oteborg University, SE-412
96 G\"oteborg, Sweden}
\author{V.M. Vinokur} \affiliation{Material Science Division,
Argonne National Laboratory, Argonne, Illinois 60439}

\date{\today}
\begin{abstract}
We investigate theoretically a mechanically assisted Kondo effect
and electric charge shuttling in nanoelectromechanical
single-electron transistor (NEM-SET). It is shown that the
mechanical motion of the central island (a small metallic
particle) with the spin results in the time dependent tunneling
width $\Gamma (t)$ which leads to effective increase of the Kondo
temperature. The time-dependent oscillating Kondo temperature
$T_K(t)$ changes the scaling behavior of the differential
conductance resulting in the suppression of transport in a strong
coupling- and its enhancement in a weak coupling regimes. The
conditions for fine-tuning of the Abrikosov-Suhl resonance and
possible experimental realization of the Kondo shuttling are
discussed.
\end{abstract}
\maketitle

Kondo resonance tunneling predicted in \cite{GR} and observed
experimentally \cite{gogo} attracts a great deal of current
attention as a possible base for manipulating spin transport.
Nanomechanical shuttling (NMS) \cite{rob1} offers a unique
platform for design of a single electron transistor (SET) in which
spin switch/transfer can be controlled electromechanically; the
first successful experimental implementation of the NMS were
reported in \cite{Kot2004,erbe}. In this Letter we develop a
theory of a nanomechanical shuttling device that utilizes Kondo
resonance effect (KR) and thus breaks the ground for a new class
of effects integrating both phenomena.

The Kondo effect in electron tunneling results from the spin
exchange between electrons in the leads and the island (quantum
dot) that couples the leads and manifests itself as a sharp zero
bias anomaly in low temperatures tunneling conductance.
Many-particle interactions and the tunneling renormalize the
electron spectrum enabling KR both for odd \cite{gogo} and even
\cite{Pust00,Kikoin01} electron occupations. In the latter case
the KR is caused by the singlet-triplet crossover in the ground
state (see \cite{glasko} for a review).

A general shuttle mechanism for a charge transfer described in
\cite{rob1} implies periodic charging and de-charging of the
oscillating nano-particle.  As the bias exceeds some threshold
value, the shuttling particle oscillates with the constant
amplitude along a classical trajectory.  A model for a
nano-electromechanical single electron transistor (NEM-SET), where
either a small (nano-scale) metallic particle or a single molecule
oscillate under the external time-dependent electric field was
studied theoretically in \cite{bo}-\cite{rob4}. In experimental
realizations, the moving dot was mounted as a electromechanical
pendulum formed by a gold clapper \cite{erbe}, or a silicon
nanopillar \cite{Kot2004}. Single electron tunneling in a
molecular conductor with the center of mass motion was realized in
\cite{park,dg05}.  The vibration induced Kondo effect in
metal-organic complexes was further explored in \cite{kkw05}.
Experimental realization of Kondo effect in electron shuttling is
being one of the most challenging tasks of current nano-physics.

In this Letter we investigate the effect of the spin degrees of
freedom on the single electron transport through the NEM-SET. We
show that mechanical shuttling of a nano-particle between the
leads causes the sequential in time reconstruction of its tunnel
electronic states and gives rise to Kondo effect.  By the analogy
with the effect of the sequential in time recharging of the
particle in ordinary charge shuttling effect \cite{rob1} we call
this phenomenon {\it Kondo shuttling}.

Building on the analogy with shuttling experiments of
\cite{erbe,Kot2004}, we consider the device where an isolated
nanomachined island oscillates between two electrodes. We,
however, are interested in a regime where the applied voltage is
low enough so that the field emission of many electrons, which was
the main mechanism of tunneling in those experiments, should be
neglected. Note further that the characteristic de Broglie wave
length associated with the dot should be much shorter than typical
displacements allowing thus for a classical treatment of the
mechanical motion of the nano-particle.   The condition
$\hbar$$\Omega$$\ll$$T_K$, necessary to eliminate decoherence
effects, requires for e.g. planar quantum dots with the Kondo
temperature $T_K$$\gtrsim$$100mK$, the condition
$\Omega$$\lesssim$$1GHz$ for oscillation frequencies to hold; this
frequency range is experimentally feasible \cite{erbe,Kot2004}.
The shuttling island then is to be considered as a "mobile quantum
impurity", and transport experiments will detect the influence of
mechanical motion on a differential conductance. If the dot is
small enough, then the Coulomb blockade guarantees the single
electron tunneling or cotunneling regime, which is necessary for
realization of Kondo effect \cite{GR,glasko}.  Cotunneling process
is accompanied by the change of spin projection in the process of
charging/discharging of the shuttle and therefore is closely
related to the spin/charge pumping problem \cite{brouw98}.
\begin{figure}[h]
  \includegraphics[width=7cm,angle=0]{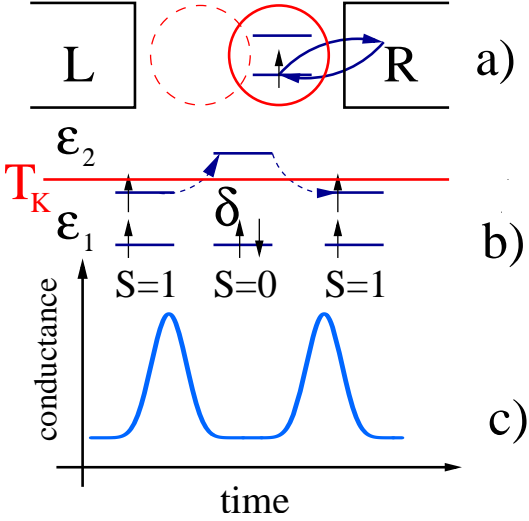}
  \caption{a) Nanomechanical resonator with the odd number of electrons as a "mobile quantum
  impurity".  b) Time evolution of the level spacing for the shuttle with the even number of electrons measured
  with respect to evolving  Kondo temperature.
  c) Pulse ZBA in tunneling conduction for Kondo shuttling with the singlet/triplet transition.
  }\label{fig:sh1}
\end{figure}

We apply our theory to planar quantum dots in
semiconductor heterostructures \cite{note1}. In these systems the
KR tunneling may be realized both for odd and even electron
occupation $N$ \cite{Kikoin01,glasko}. If $N$ is odd, the last
occupied level in the island is occupied by single electron
(Fig.1.a).  In this case the Kondo effect occurs and the
corresponding Kondo temperature, $T^0_{K}$, is that of a static
regime. The question we wish to address is how $T_K$ is {\it
influenced} by the adiabatic mechanical motion of the island. If,
on the contrary, $N$ is even, the ground state of the
nano-particle is singlet $S$$=$$0$ whereas the excited state is
triplet $S$$=$$1$ (Fig.1.b). Under the assumption that the energy
difference between the singlet and triplet states
$\Delta_{ST}$$=$$\delta$$-$$J_{ex}$ exceeds the actual Kondo
temperature  $\Delta_{ST}$$>$$T_{K}$, the Kondo effect is absent
in the static limit (here $\delta$$=$$\epsilon_2$$-$$\epsilon_1$
and $J_{ex}$ is ferromagnetic exchange, $\epsilon_{i=1,2}$ are the
energies of single-electron states for the last occupied and first
empty levels).  However the {\it mechanically induced} Kondo
effect may arise provided the inequality
$\Delta_{ST}$$<$$T_{K}(t)$ is satisfied due to mechanical motion
of the dot.

Now we turn to a quantitative description of Kondo shuttling. The
Hamiltonian $H$$=$$H_0$$+$$H_{tun}$ is given by
\begin{eqnarray}
&&H_0 =\sum_{k,\alpha}
\varepsilon_{k\sigma,\alpha}c^\dagger_{k\sigma,\alpha}c_{k\sigma,\alpha}
+\sum_{i\sigma}[\epsilon_{i}-e{\cal E}x] d^{\dagger}_{i\sigma}
d_{i\sigma} + U n^2 \nonumber
\\
&& H_{tun}=
\sum_{ik\sigma,\alpha}T^{(i)}_{\alpha}(x)[c^\dagger_{k\sigma,\alpha}d_{i\sigma}
+ H.c],\label{ham1}
\end{eqnarray}
where $c^\dagger_{k\sigma}$, $d^\dag_{i\sigma}$
create an electron in the lead $\alpha$$=$$L$$,$$R$, or the dot
level $\varepsilon_{i=1,2}$, respectively,
 $n$$=$$\sum_{i\sigma}$$d^\dagger_{i\sigma}$$
d_{i\sigma}$, ${\cal E}$ is the electric field between the leads.
The tunnelling matrix element
$T^{(i)}_{L,R}(x)$$=T^{(i,0)}_{L,R}$$\exp$$[$$\mp$$x(t)$$/$$\lambda_0]$,
depends exponentially on the ratio of the time-dependent
displacement $x(t)$ (which is considered to be a given harmonic
function of the time) and the electronic tunnelling length
$\lambda_0$.

We begin with the discussion of an odd $N$,$~S=1/2$, case
(Fig.1.a). Then only the state with $i$$=$$1$ retains in
(\ref{ham1}), and hereafter we omit this index. In order to find
an analytic solution, we assume that if $x(t)$ varies
adiabatically slow (on the scale of the tunneling recharging
time), there is no charge shuttling due to multiple recharging
processes \cite{rob1}, but the KR cotunneling occurs. The
time-dependent tunneling width is $ \Gamma_{\alpha}(t)=2\pi\rho_0
|T_\alpha(x(t))|^2$ \cite{GKN}, where $\rho_0$ is the density of
states at the leads Fermi level. The adiabaticity condition reads:
$\hbar \Omega$$\ll$$T_K\ll$$\Gamma$, with
$\Gamma$$=$$\min$$[$$\sqrt{\Gamma_L^2(t)+\Gamma_R^2(t)}$$]$. We
apply the time-dependent Schrieffer-Wolff transformation and
obtain the time-dependent Kondo Hamiltonian \cite{GKN} as
\begin{equation}
H=H_0+ \sum_{k\alpha\sigma, k'\alpha'\sigma'}{\cal
J}_{\alpha\alpha'}(t)[\vec{\sigma}_{\sigma\sigma'}\vec{S}+
\frac{1}{4}\delta_{\sigma\sigma'}]c^\dagger_{k\sigma,\alpha}c_{k'\sigma',\alpha'}
\label{ham2}
\end{equation}
where ${\cal J}_{\alpha,\alpha'}(t)$$=$$\sqrt{\Gamma_\alpha(t)
\Gamma_{\alpha'}(t)}$$/$$(\pi\rho_0 E_d(t))$ and $\vec
S$$=$$\frac{1}{2}$$d_\sigma^\dagger$$\vec\sigma_{\sigma\sigma'}$$d_{\sigma'}$.
Without the loss of the generality, we can restrict ourselves to
symmetric Anderson model ($\epsilon_0$$=$$-$$U$$/$$2$ in the
static limit) and neglect the renormalization of a single electron
level position by tunneling ($\Gamma$$/$$U$$\ll$$1$) and also by
its shift from the equilibrium position [$(e{\cal
E}A)$$/$$U$$\ll$$1$] such as $E_d(t)$$\approx$$
E_d^0$$=$$U$$/$$4$.

As long as the nano-particle does not subject to the external
time-dependent electric field, the Kondo temperature is given by
$T_K^0$$=$$D_0$$\exp$$\left[-(\pi U)/(8\Gamma_0)\right]$ (for
simplicity we assumed that
$\Gamma_L(0)$$=$$\Gamma_R(0)$$=$$\Gamma_0$;
$D_0$$=$$\sqrt{2\Gamma_0 U/\pi}$ plays the role of effective
bandwidth). As the nano-particle moves adiabatically, $\hbar
\Omega$$\ll$$\Gamma$, the decoherence effects are small provided
$\hbar$$\Omega$$\ll$$T_K^0$ (see the discussion below). In this
case the time can be treated as an external parameter, and the
renormalization group equations for the Hamiltonian (\ref{ham2})
can be solved the same manner as those for the equilibrium
\cite{GKN}. As a result, the Kondo temperature becomes time
oscillating:
\begin{eqnarray}
T_K(t)=D(t) \exp\left[-\frac{\pi
U}{8\Gamma_0\cosh(2x(t)/\lambda_0)}\right].\label{ftk}
\end{eqnarray}
Neglecting the weak time-dependence of the effective bandwidth
$D(t)$$\approx$$D_0$, we arrive at the following expression for
the time-averaged Kondo temperature:
\begin{eqnarray}
\langle T_K\rangle=T_K^0 \bigg\langle \exp\left[\frac{\pi
U}{4\Gamma_0}\frac{\sinh^2 (x(t)/\lambda_0)}{1+2\sinh^2
(x(t)/\lambda_0)}\right]\bigg\rangle. \label{tkon1}
\end{eqnarray}
Here $\langle$$...$$\rangle$ denotes averaging over the period of
the mechanical oscillation. The expression (\ref{tkon1}) acquires
especially transparent form when the amplitude of mechanical
vibrations $A$ is small: $A$$\lesssim$$\lambda_0$. In this case
the Kondo temperature can be written as
$\langle$$T_K$$\rangle$$=$$T_K^0$$\exp(-2W)$, with the
Debye-Waller-like exponent $W$$=$$-$$\pi$$ U$$ \langle$$
x^2(t)$$\rangle)$$/$$(8\Gamma_0\lambda_0^2)$, giving rise to the
enhancement of the static Kondo temperature. This
counter-intuitive exponentially {\it large} Debye-Waller factor
results from the strong asymmetry in the tunneling rate at the
turning points of the nano-particle trajectory.  Note, that taking
the formal limit of the large-amplitude oscillations
$A$$\gg$$\lambda_0$ one obtains $\langle$$
T_K$$\rangle$$\to$$T_K^{max}(A)$$\gg $$T_K^0$ for  the
time-averaged Kondo temperature if $\Gamma^{max}_{L,R}$$\ll$$U$
(see the insert in Fig.2). In the opposite limit $
\Gamma_{L,R}$$($$t_1$$<$$t$$<$$t_2$$)$$\gtrsim$$U$ the system
falls into a mixed valence regime where the Kondo temperature is
the poorly-defined quantity. The result (\ref{tkon1}) survives as
long as the large amplitude limit holds provided that the
condition $\langle$$T_K$$\rangle$$/D_0
$$\ll$$ 1$ is still fulfilled.
We conclude that the Kondo temperature considerably increases as
compared to $T_K^0$ when the shuttling particle approaches to one
of the leads. The relative variation of the Kondo temperature
oscillations at small shuttling amplitudes is given by
\begin{eqnarray}
\frac{\delta T_K}{T_K^0}=\frac{\langle T_K\rangle -
T_K^0}{T_K^0}=2\frac{\langle x^2(t)\rangle}{\lambda_D^2},
\label{tk1}
\end{eqnarray}
where
$\lambda_D$$=$$\lambda_0$$/$$\sqrt{\ln(D_0/T_K^0)}$$\ll$$\lambda_0$
is the effective tunneling length which accounts for the Kondo
renormalizations.
\begin{figure}[h]
  \includegraphics[width=60mm,angle=0]{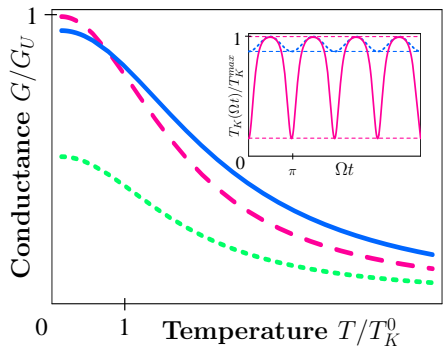}
  \caption{Differential conductance of a Kondo shuttle. Solid line denotes $G$ for the shuttle, dashed line stands
  the for static island located in
  the symmetric (central) position, dot line stands for the static island located asymmetrically.
  Insert shows the  time oscillations of $T_K$ for small (dot line) and large (solid line) shuttling amplitudes.
  }\label{fig:sh2}
\end{figure}

\vspace*{-3mm} Let us discuss the temperature behavior of the
differential conductance $G(T,V_{dc}\to 0)$$=$$dI$$/$$dV_{dc}$. In
the strong coupling Kondo limit $T$$\ll$$T_K^0$
\begin{eqnarray}
 G
=\frac{2e^2}{h}\Bigg\langle\frac{4\Gamma_L(t)\Gamma_R(t)}{(\Gamma_L(t)+\Gamma_R(t))^2}\Bigg\rangle=
\frac{2e^2}{h}\left(1-2 \frac{\delta
T_K}{T_K^0}\frac{\lambda_D^2}{\lambda_0^2}\right). \label{gk1}
\end{eqnarray}
The conductance does not reach the unitary limit $G_U=2e^2/h$ due
to the asymmetry in the respective couplings to the leads. If,
however, the shuttling island was not centrally positioned when
starting its motion, the effective magnitude of the conductance
can grow as compared to its value at the starting position (see
Fig.2).

In the weak coupling regime $T_K^{max}$$\ll$$T$$\ll$$D_0$ the zero
bias anomaly (ZBA) in the tunneling conductance is given by
\begin{eqnarray}
G(T)
=\frac{3\pi^2}{16}G_U\Bigg\langle\frac{4\Gamma_L(t)\Gamma_R(t)}
{(\Gamma_L(t)+\Gamma_R(t))^2}\frac{1}{[\ln(T/T_K(t))]^2}\Bigg\rangle.
\end{eqnarray}
Although the central position of the island is most favorable for
the Breit-Wigner (BW) resonance ($\Gamma_L$$=$$\Gamma_R$), it
corresponds to the minimal width of the Abrikosov-Suhl resonance.
The turning points  correspond to the maximum of the Kondo
temperature given by the equation (\ref{ftk}) while the system is
away from the BW resonance.  These two competing effects lead to
the effective enhancement of $G$ at high temperatures:
\begin{eqnarray}
G(T)=G^0_K\bigg\langle\left(\frac{1}{ 1-
2\alpha^2(T)\sinh^2(x(t)/\lambda_0)}\right)^2\bigg\rangle,
\label{fcon}
\end{eqnarray}
where
$G^0_K$$=$$G_U$$($$3$$\pi^2$$)$$/$$($$16$$\ln^2$$($$T$$/$$T_K^0$$)$$)$
is the conductance of the static island in the central position
(Fig.2), $\alpha^2(T)$$=$$\ln(D_0/T)$$/$$\ln(T/T_K^0)$ and
$\lambda_T$$=$$\lambda_0$$/$$\alpha(T)$ is the temperature
dependent tunneling length. Evaluating (\ref{fcon}) for the small
amplitude limit under the condition $\lambda_T$$\ll$$\lambda_0$,
we obtain
\begin{eqnarray}
\frac{\delta G_K}{G^0_K}=\frac{G(T) -G^0_K}{G^0_K}=2\frac{\delta
T_K}{T_K^0}\frac{1}{\ln(T/T_K^0)}. \label{dg1}
\end{eqnarray}
Formally, the correction to the conductance $\delta$$G_K$
(\ref{dg1}) must be compared with the regular term
$O$$($$C$$/$$[$$\ln^3$$($$T$$/$$T_K^0$$)$$]$$)$. The latter,
however has much smaller amplitude
$C$$\sim$$\ln$$($$\ln$$($$T$$/$$T_K^0$$)$$)$$\ll$$\ln$$($$D_0$$/$$T_K^0$$)$.
Thus, the Eq.(\ref{dg1}) describes the leading correction to
conductance.

In the limit $T_K^0$$\ll$$\hbar$$\Omega$$\ll$$\Gamma$ the
differential conductance in the weak coupling regime is given by
\begin{eqnarray}
G_{peak}= \frac{3\pi^2}{16}G_U \frac{1}{[\ln(\hbar/(\tau
T_K^0))]^2},\label{dc2}
\end{eqnarray}
where $\hbar/\tau\sim \hbar \Omega$ is determined by the
decoherence effects associated with the non-adiabaticity of the
motion of the shuttling-particle and by the Q-factor of the NEM
device. In general, the behavior of the differential conductance
at low temperatures has a form
$G_{peak}$$/$$G_U$$=$$F$$\left[\right.$$($$\delta$$
T_K$$/$$T_K^0$$)$$\cdot$$f$$\left(\right.$$\hbar$$\Omega$$/$$T_K^0$$\left.\right)$$\left.\right]$,
where $F(x)$ and $f(y)$ are two universal functions, each of them
depending on one variable similar to \cite{GKN}. In the large
voltage limit $eV$$\gg$$T_K^0$ the finite current transferred by
the shuttle and therefore the noise created by this current leads
to $\hbar$$/$$\tau$$\sim$$eV$ destroying the Kondo effect. We will
present more detailed discussion on the decoherence effects in the
"anti-adiabatic" Kondo regime elsewhere.

Next we turn to the case of even $N$ in the island (Fig.
\ref{fig:sh1}.b). In this case one may refer to the {\it
excited-state Kondo features} \cite{glasko,hosch}, where the KR
tunneling is possible only during the time intervals where
\begin{eqnarray}\label{beat}
\Delta_{ST}(t) = \delta(t) - J_{ex}(t) < T_K(t).\label{ineq}
\end{eqnarray}
The level spacing $\delta(t)$$=$$\epsilon_2(t)$$-$$\epsilon_1(t)$
may reduce due to the tunneling-induced Friedel shift
\begin{eqnarray}
  \epsilon_i(t) = \epsilon_i^0-\sum_{\alpha=L,R}|T^{(i)}_\alpha(t)|^2 Re
  \int \frac{\rho_0 d\varepsilon}{\epsilon_i - \epsilon_\alpha},
\end{eqnarray}
provided $T^{(2)}_\alpha$$>$$T^{(1)}_\alpha$, which is usually the
case \cite{glasko}. This effect is maximal near the turning points
of shuttle motion. Similar 2-nd order tunnel processes results in
the so called Haldane renormalization \cite{hald} of $J_{ex}$. In
close analogy with \cite{Kikoin01}, it is easy to see that the
reduction of exchange gap in a dot with even occupation obeys the
renormalization group flow equation
\begin{eqnarray}
d \Delta_{ST}/d\eta = \rho_0\sum_\alpha\left[|T^{(2)}_\alpha|^2 -
| T^{(1)}_\alpha|^2\right],
\end{eqnarray}
where $\eta$$=$$\ln$$($$D_0$$/$$D$$)$ is the scaling variable
describing the reduction of the energy scale $D$ of the band
electrons in the leads. Additional contribution to this reduction
originates from the mixture of the exited state with two electrons
on the level $\varepsilon_2$ with the ground state singlet
\cite{Kikoin01}. These effects are also maximal around the turning
points of the shuttle trajectory.

Thus, if the condition (\ref{beat}) is valid for the certain time
intervals during the oscillation cycle (Fig.1.b), the Kondo
tunneling is possible  for a part of this cycle, where the shuttle
is close to one of the leads. It should be emphasized that in this
regime only the weak-coupling Kondo effect may be observed at $T$$
\gg$$T_K$, whereas at $T$$\to$$0$ the triplet state is quenched
and the dot behaves as a zero spin nano-particle \cite{hosch}. The
full scale Kondo effect may arise only if the variation of $|
T^{(i)}_\alpha(t)|^2$ induces the crossover from a singlet to a
triplet ground state of a shuttle. The singlet/triplet crossover
induced by the variation of gate voltages was observed on a static
planar dot \cite{kogan}. Since the adiabaticity condition
$\hbar$$\Omega$$\ll T_K^{(S=1)}$ is violated close to the
singlet/triplet transition, our approach does not apply to this
regime.

In excited-state Kondo regime $T_K$ depends on the value of
$\Delta_{ST}$, being scaled relative to the true $S$$=$$1$ value
$T^{(S=1)}_{K}$ in accordance with the law
$T_K(t)$$/$$T^{(S=1)}_{K}$$=$$\left[\right.$$T^{(S=1)}_{K}$$/$$\Delta_{ST}(t)$$\left.\right]^\zeta$
with $\zeta$$\lesssim$$1$ \cite{Kikoin01}. Thus, we conclude that
the Kondo shuttling in case of even $N$ may be observed as a {\it
pulse ZBA in tunnel conduction} (Fig.1.c), which emerges in time
intervals $\delta$$t$, where the condition (\ref{beat}) is
fulfilled. Assuming the linear dependence $\Delta_{ST}$$($$x$$)$
\cite{kogan}, we estimate these intervals as $\delta
t$$\sim$$\delta x$$/$$($$\Omega$$\sqrt{\langle x^2}$$\rangle$$)$,
where $\delta$$x$ is the distance from the turning point at which
the Kondo shuttling is possible \cite{ftb}.

The Kondo shuttling differs from the  Kondo effect in molecular
conductor with the centrum of mass motion \cite{dg05}. In that
case the shift of the centrum of mass is caused by single electron
transport, and the {\it non-adiabatic} phonon-assisted processes
interfere with the Kondo tunneling.

In conclusion, we have found that the Kondo shuttling in NEM-SET
increases the Kondo temperature due to the asymmetry of coupling
in the turning points compared to central position of the island.
As a result, in the case of odd $N$ the differential conductance
is enhanced in the weak coupling regime and is suppressed in the
strong coupling limit. In the case of even $N$, Kondo tunneling
exists only as a shuttling effect.

We acknowledge useful discussions with K. Matveev and L.W.
Molenkamp. MK and RS are grateful to ANL for the hospitality
during their visit. Research in Argonne was supported by U.S. DOE,
Office of Science, under Contract No. W-31-109-ENG-39. KK is
supported by ISF grant, MK acknowledges support through the
Heisenberg program of the DFG. RS is supported by EU (CANEL) and
Swedish (VR and SSF) grants.


\vspace*{-5mm}
\begin{thebibliography}{99}
\bibitem{GR} L.I. Glazman and M.E. Raikh, Pis'ma Zh. Eksp. Theor.
Fiz. {\bf 47}, 378 (1988) [JETP Lett. {\bf 47}, 452 (1988)];
T.K. Ng and P.A. Lee, Phys. Rev. Lett. {\bf 61}, 1768 (1988).


\bibitem{gogo} D.Goldhaber-Gordon {\it et at}, Nature (London)
{\bf 391}, 156 (1998), S.M.Cronenwett {\it et al}, Science {\bf
281}, 540 (1998).

\bibitem{rob1} L.Y.Gorelik {\it et al},
Phys. Rev. Lett. {\bf 80}, 4526 (1998).



\bibitem{Kot2004} D.V. Scheible {\it et al},
Phys.Rev.Lett.{\bf 93}, 186801 (2004), D.V.Scheible, R.H.Blick,
Appl.Phys.Lett.{\bf 84},4632 (2004).

\bibitem{erbe} A.Erbe {\it et al}, Phys. Rev. Lett. {\bf 87}
096106 (2001).

\bibitem{Pust00} M. Pustilnik {\it et al},
Phys. Rev. Lett. {\bf 84}, 1756 (2000).
\bibitem{Kikoin01} K. Kikoin and Y. Avishai,
Phys. Rev. Lett. {\bf 86}, 2090 (2001); Phys.  Rev.  B  {\bf 65},
115329 (2002).


\bibitem{glasko} M. Pustilnik  and L.I. Glazman, J. Phys.: Condens. Mat. {\bf16}, R513 (2004).



\bibitem{bo} D.Boese and H.Schoeller, Europhys. Lett. {\bf 54}, 668
(2001).

\bibitem{rob2} T. Nord {\it et al},
Phys. Rev. B {\bf 65}, 165312 (2002)

\bibitem{rob5} L.Y. Gorelik {\it et al},
Phys. Rev. Lett. {\bf 91}, 088301 (2003).

\bibitem{nov} T. Novotny {\it et al},
Phys. Rev. Lett. {\bf 90}, 256801 (2003).

\bibitem{rob3} D. Fedorets
{\it et al},
Phys. Rev. Lett. {\bf 92}, 166801 (2004).

\bibitem{rob4} L.Y. Gorelik {\it et al},
Phys. Rev. Lett. {\bf 95}, 116806 (2005).

\bibitem{park} H.Park {\it et al}, Nature (London) {\bf 407}, 57 (2000).

\bibitem{dg05} K.A.Al-Hassanieh {\it et al},
Phys.Rev.Lett.{\bf 95},256807 (2005).
\bibitem{kkw05} K.Kikoin {\it et al},
cond-mat/0511369.

\bibitem{brouw98} P.W.Brouwer, Phys.Rev. B {\bf 58}, R10135
(1998);
P.Sharma and P.W.Brouwer, Phys. Rev. Lett. {\bf 91}, 166801
(2003);
T.Aono, Phys. Rev. Lett. {\bf 93}, 116601 (2004).

\bibitem{note1} The GaAs/GaAlAs heterostructures can serve as an
experimental realization.

\bibitem{GKN} A.Kaminski,
Yu.V.Nazarov, and L.I.Glazman, Phys. Rev. Lett. {\bf 83}, 384
(1999); Phys. Rev. B {\bf 62}, 8154 (2000).

\bibitem{hosch} W. Hofstetter and H. Schoeller, Phys. Rev. Lett.
{\bf88}, 016803 (2002).

\bibitem{hald} F.D.M. Haldane, Phys. Rev. Lett. {\bf40}, 416 (1978).

\bibitem{kogan} A. Kogan {\it et al}, Phys. Rev. B {\bf67}, 113309
(2003).

\bibitem{ftb} The Kondo shuttling is sensitive to the external magnetic
field due to the significant Larmor shift of $\Delta_{ST}$ in
planar quantum dots.
\end{thebibliography}
\end{document}